\documentclass[letter]{jpsj2} 
\title{Coherence of Spin-Polarized Fermions Interacting with a Clock Laser\\ in a Stark-Shift-Free Optical Lattice}

\author{Masao \textsc{Takamoto}$^{1,2}$ and Hidetoshi \textsc{Katori}$^{1,2}$\thanks{E-mail: katori@amo.t.u-tokyo.ac.jp}}

\inst{$^{1}$Department of Applied Physics, Graduate School of Engineering, The University of Tokyo,
Bunkyo-ku, 113-8656 Tokyo, Japan\\
$^{2}$CREST, Japan Science and Technology Agency, 4-1-8 Honcho Kawaguchi, 332-0012 Saitama, Japan}

\abst{We investigated the coherence of spin-polarized $^{87}$Sr atoms trapped in a light-shift-free one-dimensional optical lattice during their interaction with a clock laser on the $^1S_0-{}^3P_0$ transition.  
Collapses and revivals appeared for more than 50 Rabi cycles,  attributed to the thermal distribution of discrete vibrational states in the lattice potential. The population oscillation in the clock states lasted more than 1 s, demonstrating high immunity from decoherence. This long atomic coherence suggests the feasibility of Pauli blocking of collisions in optical clock excitation.}

\kword{atomic coherence, spin polarization, Fermi statistics, cold collisions, Rabi oscillation, collapses and revivals, optical lattice, atomic clock}

\begin{document}
\maketitle

Unprecedented controllability over system parameters makes ultracold atoms in optical lattices a powerful tool for investigating quantum computations \cite{Por07} and quantum simulations of strongly correlated systems \cite{Jak98,Jak05}.
Particular attention has been paid to fermionic atoms confined in optical lattices, as they are expected to simulate the Hubbard Hamiltonian that conjectures high-temperature superconductivity in cuprates \cite{Hof02}. 
Moreover,  the noninteracting nature of fermions originating from the Pauli principle provides the system with potential applications in qubit registers \cite{Rab03,Viv04} and a variety of precision measurement, such as interferometers \cite{Ott04} and high-performance atomic clocks \cite{Gib95}. All of these applications require a minimum atomic decoherence, preferably including atom-laser interaction processes  to allow the optical manipulation of atomic states.

In this Letter, we report the coherence properties of spin-polarized $^{87}$Sr atoms interacting with resonant light.
Atoms were trapped in a one-dimensional (1D) optical lattice operated at the ``magic wavelength,'' where the light shift perturbation is canceled out in the relevant transition \cite{Kat03}. 
The Lamb-Dicke confinement provided by the lattice potential allowed the system to be free of atomic-motion-induced decoherence.
We investigated Rabi oscillations on the $^1S_0(F=9/2)-{}^3P_0(F=9/2)$ clock transition  at 429~THz.
Collapses and revivals reminiscent of those described by the Jaynes-Cummings model \cite{Blo92,Cir94} were observed up to 50 Rabi cycles.
Furthermore, the oscillations of the atomic population lasted more than 1 s, demonstrating a remarkably long coherence of the atom-laser system.
The origins of  decoherence mechanisms were investigated experimentally and numerically.
The observed long atomic coherence 
suggests that the Pauli blocking of collisions is feasible for 1D optical lattice clocks with spin-polarized fermions \cite{Tak05}. 

In precision spectroscopy, better signal-to-noise ratios are obtained by increasing the number of atoms, hence atomic density. However, this is often encountered with correspondingly larger atomic interactions that result in problematic systematic uncertainties.
Thus, some state-of-the-art atomic clocks such as  Cs fountain clocks \cite{Per02} and ballistic Ca clocks \cite{Wil02}, which use ultracold bosonic atoms, suffer from collisional frequency shifts.
An ultracold fermionic system is expected to be  advantageous in such applications, since the mean field energy due to $s$-wave interaction disappears \cite{Dem01} as a consequence of the Pauli principle, and $p$-wave interactions are suppressed at ultralow temperatures.
The application of fermionic $^{134}$Cs was suggested to eliminate the cold collision shift of Cs microwave clocks \cite{Gib95}. 
The disappearance of the mean field energy was demonstrated experimentally  in an RF transition with quantum degenerate $^6$Li atoms held in a dipole trap \cite{Gup03}.

\begin{figure}[t]
\begin{center}
\includegraphics[width=0.9\linewidth]{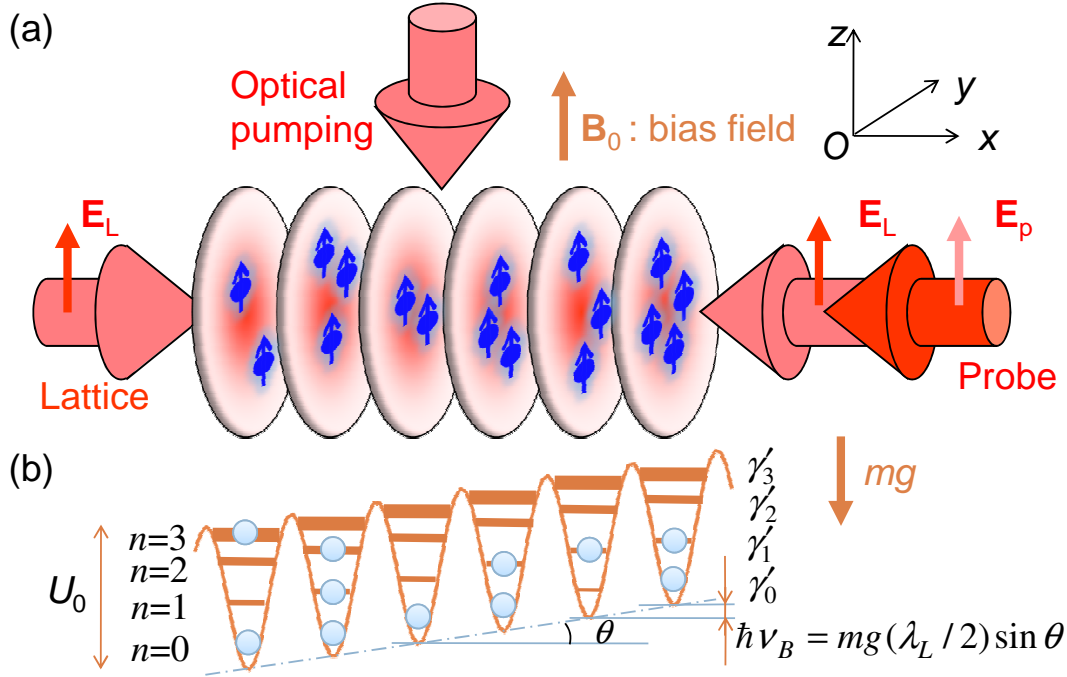}
\caption{(Color online) (a) Ultracold $^{87}$Sr atoms with a temperature of a few $\mu$K  were loaded into a one-dimensional (1D) optical lattice operated at the magic wavelength of $\lambda_L=813.42~{\rm nm}$.
A circularly polarized laser optically pumped  atoms into the $^1S_0(F=m_F=9/2)$ stretched state. 
(b) The  energies of atoms in a 1D lattice show the Bloch bands.
This bandwidth can be reduced by tilting the lattice $\theta$ with respect to the horizontal plane to introduce the gravitational potential difference $\hbar\nu_B$ between adjacent sites.}
\label{setup} 
\end{center}
\end{figure}

To apply  the Pauli blocking of collisions in atomic clocks, the fermionic system needs to be uniformly excited so as to preserve  fermionic identity.
At an RF wavelength, as the spatial spread $\Delta x$ of laser-cooled  atoms  can be much smaller than the relevant transition wavelength $\lambda_0$, the Lamb-Dicke condition $\Delta x \ll \lambda_0$ is well satisfied; therefore, the system is  uniformly excited without being significantly affected by  Doppler shifts. 
However, its application to optical transitions remains technically challenging \cite{Tak06,Lud08}, because the much shorter transition wavelength $\lambda_0$ of less than a $\mu$m makes the Lamb-Dicke condition far more severe. 
We show that optical lattices operated at the magic wavelength offer an ideal tool for investigating such possibilities.

We consider a system where spin-polarized fermionic atoms are trapped in an array of disk-shaped potentials (1D optical lattice) formed by a standing wave of a Gaussian beam, as shown in Fig.~\ref{setup}. 
Atoms are excited on the $\left| S\right\rangle  \otimes \left| {\bf n} \right\rangle  \to \left| P \right\rangle  \otimes \left| {\bf n} \right\rangle$ electronic-vibrational transition by a probe laser with a wave vector ${\bf k}_p=(k_x,k_y,k_z)$.  
For an atom at a position ${\bf r}=(x,y,z)$ in the vibrational state ${\bf n}=(n_x,n_y,n_z)$ with trapping frequencies $(\nu_x,\nu_y,\nu_z)$, the effective Rabi frequency for the carrier (${\bf n}\rightarrow {\bf n}$) component is given by \cite{Lei03}
\[
\Omega _{\rm eff}  = \Omega _{\rm el} \prod\limits_{j = x,y,z} \left| {\left\langle {n_j } \right|\exp (ik_j j)\left| {n_j } \right\rangle } \right|,
\]
where $\Omega_{\rm el}$ is the Rabi frequency for the electronic transition. 
When ${\bf k}_p$ is nearly parallel to ${\bf k}_L(\parallel\hat{x})$ of a lattice laser so that $k_y y\approx k_z z\approx 0$ holds, the effective Rabi frequency $\Omega _{\rm eff}$ is  determined solely by the $x$-motional state $n=n_x$ as
\begin{equation}
\Omega_n\approx \Omega _{\rm el} \exp ( -  \eta _x^2 /2)L_{n}(\eta _x^2 ).
\label{Rabi}
\end{equation}
Here, $\eta _x  = k_x\sqrt {h/(2m\nu _x )} /(2\pi)$ is the Lamb-Dicke parameter with a mass $m$ and a Planck constant $h$. 
$L_n(x)$ is the $n$-th order Laguerre polynomial.
As a consequence of the $n$-dependent vibrational coupling strength, the thermal occupation over $n$ states  leads to collapses and revivals of Rabi oscillations.

The experimental setup is similar to that described elsewhere \cite{Tak06}.
Ultracold fermionic $^{87}$Sr atoms with a temperature of a few $\mu$K  \cite{Muk03} were loaded into a one-dimensional (1D) optical lattice operated at a magic wavelength $\lambda_L=813.428~{\rm nm}$ with a wave vector $\pm {\bf k}_L(\parallel \hat{x})$ and an electric field ${\bf E}_L(\parallel \hat{z})$. 
Typically, $ 10^4$ atoms were trapped in about 500 lattice sites spread over ${200 }\,\mu$m.
By focusing the lattice laser with an intensity of 100~mW into an $e^{-2}$ beam radius of $30~{\mu}$m, the trap frequencies and lattice depth were $\nu_x\approx 47$~kHz and $\nu_{y,z}\approx 300$~Hz, and $U_0\approx 45E_R$, respectively, with $E_R=(h/\lambda_L)^2/2m$ the recoil energy. 
The lattice laser was directed at an angle $0\leq \theta\leq 20$~mrad with respect to the horizontal plane.
This gave a gravitational potential difference $\hbar\nu_B = m g (\lambda_L/2) \sin \theta$ between adjacent sites, which reduced the Bloch bandwidth  $\gamma_n$ of the optical lattice to $\gamma_n'$.

A circularly polarized $(\sigma^+)$ laser resonant to the $^1S_0(F=9/2)-{}^3P_1(F=9/2)$ transition was applied along a bias magnetic field ${\bf B}_0(\parallel \hat{z})=10$~mG to optically pump atoms into the outermost $^1S_0(F=m_F=9/2)\equiv \left| S \right\rangle $ state.
Rabi oscillations were observed by exciting  atoms to the long-lived $^3P_0(F=m_F=9/2)\equiv \left| P \right\rangle $  state with a lifetime of 150~s. The probe laser was linearly polarized ${\bf E}_p(\parallel {\bf B}_0)$ with a linewidth of about 10~Hz at $\lambda_p=698$~nm.
Its wave vector ${\bf k}_p$ was parallel to ${\bf k}_L$ within 2.5~mrad 
\cite{ft1} 
to reduce the Lamb-Dicke parameter for the radial direction $\eta_{y,z}<1\times10^{-2} $. 
Consequently, the effective Rabi frequency $\Omega_n$ for the $\left| S\right\rangle  \otimes \left| n \right\rangle  \to \left| P \right\rangle  \otimes \left| n \right\rangle$ transition  is approximated using eq.~(\ref{Rabi}) with the Lamb-Dicke parameter $\eta _x \approx0.31$. Hereafter, $|n\rangle$ denotes the vibrational states of the $x$-motion.

Before observing Rabi oscillations, we applied a state purification process to achieve higher spin polarization, as well as to depopulate atoms in the  higher $n(\geq 2)$ vibrational states. First, atoms were excited on the $|S\rangle\rightarrow | P \rangle $ clock transition by a 1-ms-long probe pulse with an area of $\pi$. 
The remaining atoms in the $|S\rangle$ state, which occupied  unwanted Zeeman substates $m_F\neq9/2$ or higher vibrational states, were heated out of the lattices by shining a  resonant laser on the $^1S_0(F=9/2)-{}^1P_1(F=11/2)$ transition. Then,  atoms in the $|P \rangle$ state were pumped back to the $|S\rangle$ ground state by the same probe pulse as above.

Rabi oscillations were measured by determining the atom excitation probability as $P(\tau_R)=N_P/(N_S+N_P)$, after applying the resonant probe laser for a duration of $\tau_R$, where $N_S$ and $N_P$ were the number of atoms   in the $|S\rangle$ and $|P\rangle$ states, respectively.  
The laser-induced fluorescence (LIF) intensity $I_S=\kappa N_S$, where $\kappa$ gave photon detection efficiency per atom, was measured by shining a laser nearly resonant to the $^1S_0-{}^1P_1$ transition. 
After throwing the atoms  in the $|S\rangle$ state away, another $\pi$ probe pulse was applied to deexcite the atoms in the $|P\rangle$ state to the $|S\rangle$ state, where the LIF intensity $I_P(=\kappa N_P)$ was measured in the same manner as before.

\begin{figure}[t]
\begin{center}
\includegraphics[width=0.9\linewidth]{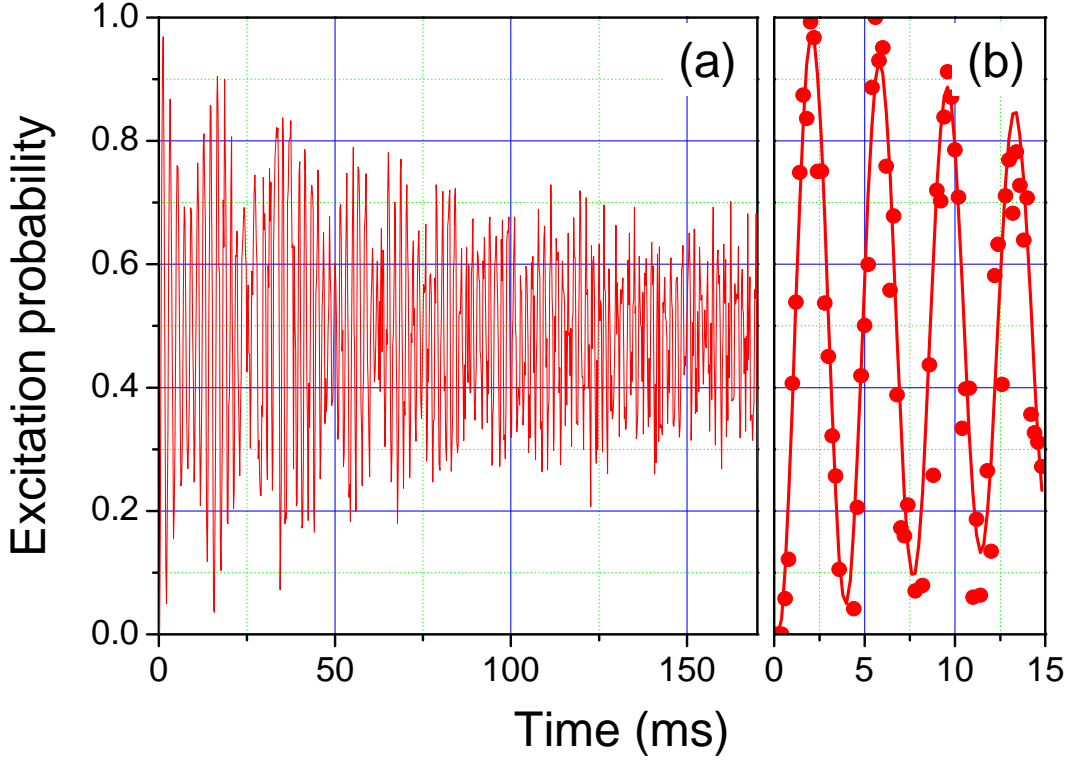}
\caption{(Color online) Rabi oscillation of atoms on the $^1S_0(F=m_F=9/2)-{}^3P_0(F=m_F=9/2)$ clock transition. (a) Collapses and revivals were observed up to 100~ms. (b) For the first Rabi excitation at 2~ms, more than 98 \% of the atoms were excited, demonstrating a high degree of spin polarization. The filled circles indicate measured points; each point required a measurement time of about 1~s.
}
\label{Rabi_exp1}
\end{center}
\end{figure}

Figure~\ref{Rabi_exp1}(a) shows $P(\tau_R)$ as a function of the duration of the probe laser with Rabi frequency $\Omega_0/2\pi\approx {500}$~Hz for atoms in the $n=0$  state.
During the Rabi measurements that took about 1~s for each data point, probe laser frequency was  stabilized to within a few Hz by alternately referencing the $\left| S \right\rangle  \to \left| P \right\rangle$ transition \cite{Tak06} every 2 s and its intensity was  stabilized to within 1 \% to keep Rabi frequency constant.
Collapses and revivals of Rabi oscillations occurred at $\sim 9 \times (2j-1)$~ms and $\sim 9\times 2j$~ms with $j=1,2,3,...$, respectively.
The envelope of the Rabi oscillations was given by the beat note $|\Omega_0-\Omega_1|$ of atoms in the $n=0$ and 1 vibrational states.
Figure 2(b) shows a close-up view of the initial oscillations at $\Omega_0/2\pi\approx { 250}$~Hz.
In the first $\pi$-pulse excitation at $\tau_R=2$~ms, more than 98 \% of the atoms were excited on the clock transition, demonstrating that the  purification process produced a  high degree of spin polarization.

\begin{figure}[t]
\begin{center}
\includegraphics[width=0.9\linewidth]{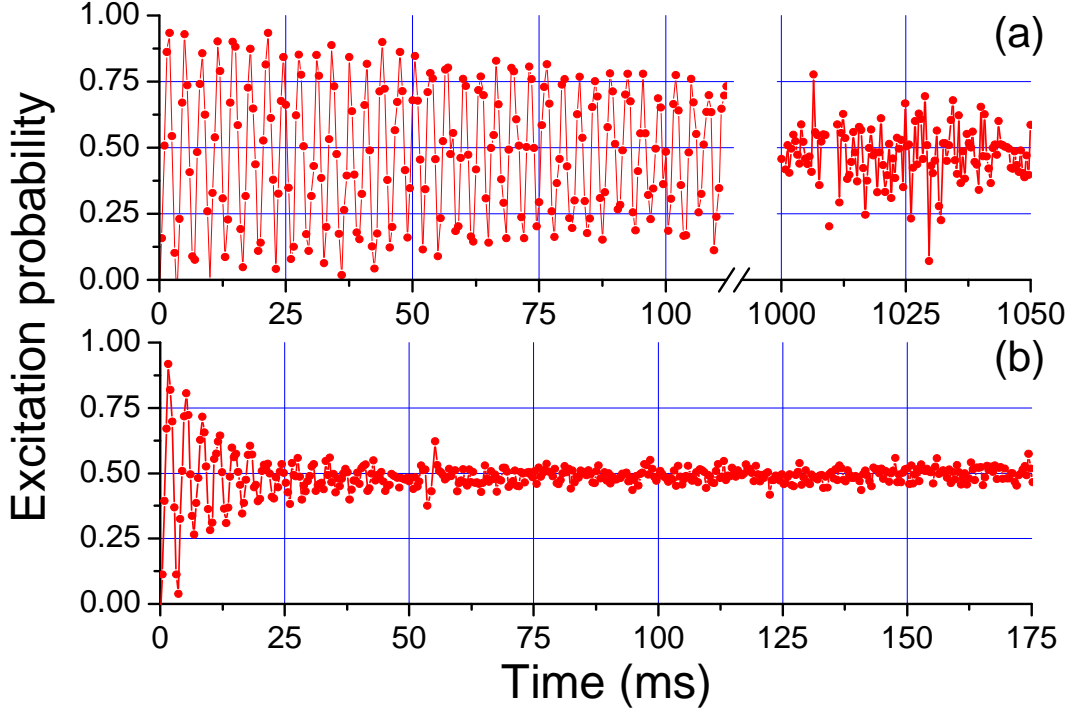}
\caption{(Color online) Amplitude decay of Rabi oscillations. (a) Although the phase of the Rabi oscillation started to jitter after a laser coherence time of about 100~ms, the population oscillation continued for more than 1~s. 
(b) By introducing  inhomogeneity in atom-laser interactions, the population oscillation decayed  rapidly and approached $P\approx 0.5$. 
}
\label{Rabi_exp2}
\end{center}
\end{figure}

Although the Rabi oscillations started to jitter for $\tau_R\geq 100$~ms, which was roughly our laser coherence time, the population oscillation lasted more than 1 s, as shown in Fig.~\ref{Rabi_exp2}(a).  
This is because  an ensemble of atoms that constituted each data point accumulated the same probe laser phase, even if it were jittery, as long as no inhomogeneous perturbations came in.
This long population oscillation demonstrated magic lattices' high immunity from decoherence.
Here, to realize nearly single-component Rabi oscillation with an $n=0$ state, the atom population at the  $n\geq 1$ vibrational levels was reduced  by gradually lowering the lattice depth down to $U_0\approx 30 E_R$ 
\cite{ft2}.
Truncated by the gravitational potential \cite{Kat99}, the radial potential depth was as small as $10E_R$, which was  crucial for selecting atoms with low radial velocities to moderate dephasing due to radial motion.

To demonstrate the dephasing by inhomogeneous perturbations, the probe laser beam was misaligned with respect to the lattice laser by about 5~mrad so as to be sensitive to the radial motion.
We observed a rapidly decaying Rabi oscillation, as shown in Fig.~\ref{Rabi_exp2}(b). 
Here, the Lamb-Dicke parameter for the radial direction increased to $\eta_{y}\approx 2 \times 10^{-2}$ and the approximation in eq.~(1) was no longer applicable, as the vibrational coupling strength $\exp ( -  \eta _{y}^2 /2)L_{n}(\eta _{y}^2 )$ severely affected each atom's phase evolution.
A much longer waiting time before revival would be expected, since radial vibrational states $k_B T/h \nu_{y}\sim 10^2$, which are far more than that discussed in Fig.~\ref{Rabi_exp1}, were contributed to the dephasing of the Rabi oscillation.
 
The observed population decay may also be affected by a finite trap lifetime, as well as  lattice photon scattering time. 
At a vacuum pressure of $5\times 10^{-9}$~Torr, the atom trapping lifetimes for the $|S\rangle$ and $|P\rangle$ states were measured to be $\tau_S=1.4$~s and $\tau_P=0.5$~s, respectively.
These trap lifetimes would be  limited by glancing collisions with background gases.
A collision loss rate nearly three times larger for the $^3P_0$ state than for the $^1S_0$ state may be responsible for extra collision channels, such as energy pooling  and fine state changing collisions in the $^3P_J$ manifolds \cite{Kel88}.
Lattice photon scattering lifetimes were estimated to be $\tau_{LS}=12$~s and $\tau_{LP}=1.1$~s for a lattice with $U_0=45E_R$.

We simulated atomic population dynamics using a density matrix $\rho(n)$ for an atom trapped in the $n$-th vibrational state of the lattice potential. 
In optical Bloch equations (OBEs), we applied  measured trap decay rates of $\tau_S^{-1}$ and $\tau_P^{-1}$ as the longitudinal decay rates for $\rho_{S\!S}(\equiv\left| S \right\rangle \left\langle S \right|)$ and $\rho_{P\!P}(\equiv\left| P \right\rangle \left\langle P \right|)$, respectively.
The transverse decay rate $\beta_n$ was given by
\[
\beta _n {\rm{ = }}\left( {\gamma_L + 1.9 \gamma _n' } \right) + \textstyle{1 \over 2}\left(\tau _S^{-1}+\tau _P^{-1}\right),
\]
where  $\gamma_L=\tau _{LS}^{-1}+\tau _{LP}^{-1}$ is the photon scattering rate for the lattice and $\gamma_n'$ is the Bloch bandwidth for the $n$-th vibrational state.  
The effective bandwidth of $ 1.9 \gamma _n'$ was derived assuming that a uniformly populated Bloch band was excited by  a probe laser momentum of $\hbar{\bf k}_p$ \cite{Lem05}.

\begin{figure}[t]
\begin{center}
\includegraphics[width=0.9\linewidth]{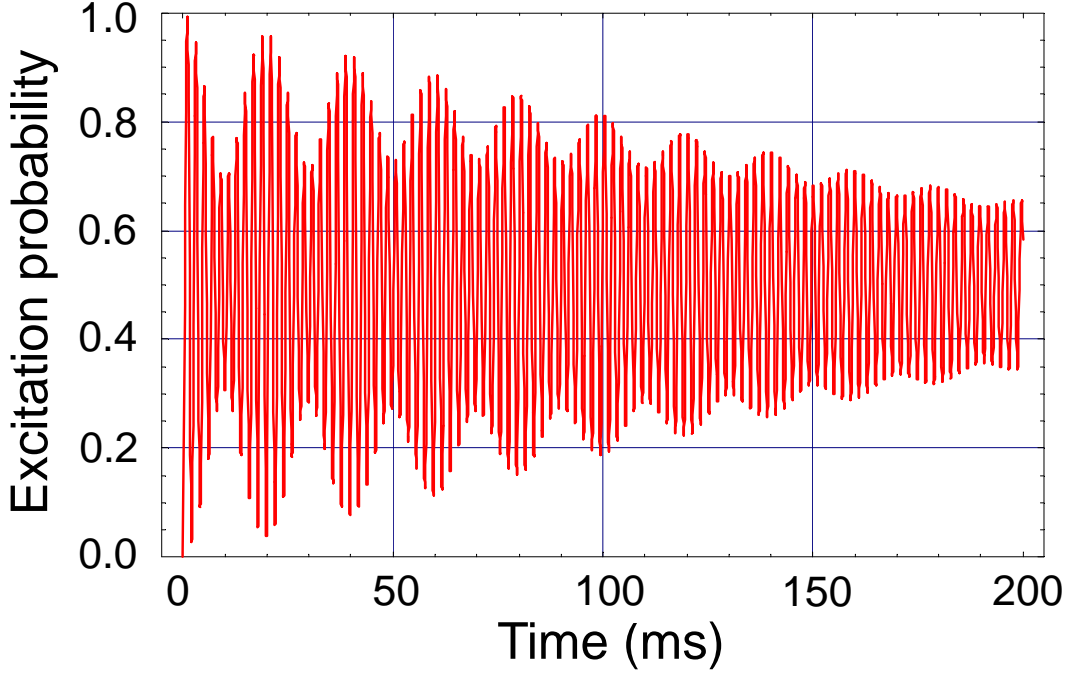}
\caption{(Color online) Excitation probability $P_{\rm OBE}(t)$ was calculated for atoms trapped in a lattice with a depth of $U_0= 45 E_R$ and thermally distributed in $n=0,\,1$ states as measured in the plot in Fig.~\ref{Rabi_exp1}(a).
}
\label{Rabi_calc}
\end{center}
\end{figure}

Figure \ref{Rabi_calc} shows $P_{\rm OBE}(\tau_R)=\left\langle \rho  \right\rangle_{P\!P}/(\left\langle \rho  \right\rangle_{S\!S}+\left\langle \rho  \right\rangle_{P\!P}$) obtained for the experimental parameters used in Fig.~\ref{Rabi_exp1}. 
The Bloch bandwidth for the $45E_R$ deep lattice was calculated to be $\gamma_0/2\pi=0.75$~Hz and $\gamma_1/2\pi=32$~Hz.
As the  tilt angle $\theta\approx  20$~mrad used in the corresponding measurement gave marginal values between the Bloch state $(\nu_B=0)$ and the Wannier-Stark state $(\nu_B> \gamma_n)$, we assumed an asymptotic bandwidth $\gamma_{n}'$ derived experimentally \cite{Sia08} as
$\gamma _{n}' \approx \gamma_n /\left[ {1 + (\nu_B /0.12\gamma_n )^2 } \right]$. 
We calculated the thermal average of the signal by $\left\langle \rho  \right\rangle  = \sum\limits_{n = 0}^{n_m} {p_{{\rm{th}}}(n) \rho (n)}$, where $p_{{\rm{th}}} (n)$ was the thermal occupation for the $n$-th vibrational state for the atomic temperature of $2~\mu$K and $n_m=1$ was the truncation of vibrational states provided by the state purification process. 
The result showed that the contrast of collapses and revivals decreased with the Bloch bandwidth $\gamma_1'$. 
Finally, Rabi oscillations with $n=0$ atoms survived. 
In this simulation, in order to fit the slow decay of the Rabi envelope, we included the influence of radial motion  by setting $\eta_{y}=5\times10^{-3}$, which reasonably agreed with our experiment.
A slight reduction in oscillation amplitude in Fig.~2(a) may be attributed to other decoherence mechanisms, such as interstate atomic collisions. 

In summary, we investigated the atomic coherence of spin-polarized fermions trapped in a 1D magic wavelength optical lattice during their interactions with a clock laser. 
The thermal population distribution among a few axial vibrational states showed collapses and revivals for more than 50 Rabi cycles.
More interestingly, by carefully removing inhomogeneous atom-laser interactions, we observed  oscillations of the atomic population for more than 1 s, which was beyond our laser coherence time.
This long atomic coherence, which was  limited by residual gas collisions and lattice photon scattering, suggested the feasibility of applying the Pauli blocking of collisions in spin-polarized 1D optical lattice clocks with fermionic atoms, e.g., $^{87}$Sr \cite{Tak06}, where relevant $p$-wave centrifugal barriers of 96~$\mu$K (74~$\mu$K) for the $^1S_0$ ($^3P_0$) state \cite{Tak06,San04} are significantly higher than the atomic temperature of a few $\mu$K. 
For a longer $\pi$-pulse up to 1~s, time axes of our results such as those shown in Fig.~2, can be  extended by simply reducing $\Omega_{\rm el}$ in Eq.~(1), which will allow nearly unit excitation for the first $\pi$ pulse.
To meet the necessary transverse relaxation time, a vacuum pressure of $10^{-11}$~Torr and a shallow lattice of $5E_R$ oriented vertically \cite{Lem05} should be used.
By applying a  Fermi degenerate atomic sample, a single-atom picture, as employed here, will no longer be valid, and the Bloch bandwidth will be quantum-mechanically reduced by the Pauli principle.
These directions will provide a new avenue for experimentally investigating atomic correlations reinforced by its unusually high energy resolution.

H.K. would like to thank K. Gibble, T. Ido, and M. Yamashita for helpful conversations.




\begin{thebibliography}{25}
\expandafter\ifx\csname natexlab\endcsname\relax\def\natexlab#1{#1}\fi
\expandafter\ifx\csname bibnamefont\endcsname\relax
  \def\bibnamefont#1{#1}\fi
\expandafter\ifx\csname bibfnamefont\endcsname\relax
  \def\bibfnamefont#1{#1}\fi
\expandafter\ifx\csname citenamefont\endcsname\relax
  \def\citenamefont#1{#1}\fi
\expandafter\ifx\csname url\endcsname\relax
  \def\url#1{\texttt{#1}}\fi
\expandafter\ifx\csname urlprefix\endcsname\relax\def\urlprefix{URL }\fi
\providecommand{\bibinfo}[2]{#2}
\providecommand{\eprint}[2][]{\url{#2}}

\bibitem{Por07}
\bibinfo{author}{\bibfnamefont{M.}~\bibnamefont{Anderlini}},
  \bibinfo{author}{\bibfnamefont{P.~J.}~\bibnamefont{Lee}},
  \bibinfo{author}{\bibfnamefont{B.~L.}~\bibnamefont{Brown}},
  \bibinfo{author}{\bibfnamefont{J.}~\bibnamefont{Sebby-Strabley}},
  \bibinfo{author}{\bibfnamefont{W.~D.}~\bibnamefont{Phillips}},
  \bibnamefont{and} \bibinfo{author}{\bibfnamefont{J.~V.}~\bibnamefont{Porto}}:
  \bibinfo{journal}{Nature}
  \textbf{\bibinfo{volume}{448}} (\bibinfo{year}{2007}) \bibinfo{pages}{452}.

\bibitem{Jak98}
\bibinfo{author}{\bibfnamefont{D.}~\bibnamefont{Jaksch}},
  \bibinfo{author}{\bibfnamefont{C.}~\bibnamefont{Bruder}},
  \bibinfo{author}{\bibfnamefont{J.~I.} \bibnamefont{Cirac}},
  \bibinfo{author}{\bibfnamefont{C.~W.} \bibnamefont{Gardiner}},
  \bibnamefont{and} \bibinfo{author}{\bibfnamefont{P.}~\bibnamefont{Zoller}}:
  \bibinfo{journal}{Phys. Rev. Lett.} \textbf{\bibinfo{volume}{81}} (\bibinfo{year}{1998})
  \bibinfo{pages}{3108}.

\bibitem{Jak05}
\bibinfo{author}{\bibfnamefont{D.}~\bibnamefont{Jaksch}} \bibnamefont{and}
  \bibinfo{author}{\bibfnamefont{P.}~\bibnamefont{Zoller}}:
  \bibinfo{journal}{Ann. Phys.} \textbf{\bibinfo{volume}{315}} (\bibinfo{year}{2005})
  \bibinfo{pages}{52}.

\bibitem{Hof02}
\bibinfo{author}{\bibfnamefont{W.}~\bibnamefont{Hofstetter}},
  \bibinfo{author}{\bibfnamefont{J.~I.} \bibnamefont{Cirac}},
  \bibinfo{author}{\bibfnamefont{P.}~\bibnamefont{Zoller}},
  \bibinfo{author}{\bibfnamefont{E.}~\bibnamefont{Demler}}, \bibnamefont{and}
  \bibinfo{author}{\bibfnamefont{M.~D.} \bibnamefont{Lukin}}:
  \bibinfo{journal}{Phys. Rev. Lett.} \textbf{\bibinfo{volume}{89}} (\bibinfo{year}{2002})
  \bibinfo{pages}{220407}.

\bibitem{Rab03}
\bibinfo{author}{\bibfnamefont{P.}~\bibnamefont{Rabl}},
  \bibinfo{author}{\bibfnamefont{A.~J.} \bibnamefont{Daley}},
  \bibinfo{author}{\bibfnamefont{P.~O.} \bibnamefont{Fedichev}},
  \bibinfo{author}{\bibfnamefont{J.~I.} \bibnamefont{Cirac}}, \bibnamefont{and}
  \bibinfo{author}{\bibfnamefont{P.}~\bibnamefont{Zoller}}:
  \bibinfo{journal}{Phys. Rev. Lett.} \textbf{\bibinfo{volume}{91}} (\bibinfo{year}{2003})
  \bibinfo{pages}{110403}.

\bibitem{Viv04}
\bibinfo{author}{\bibfnamefont{L.}~\bibnamefont{Viverit}},
  \bibinfo{author}{\bibfnamefont{C.}~\bibnamefont{Menotti}},
  \bibinfo{author}{\bibfnamefont{T.}~\bibnamefont{Calarco}}, \bibnamefont{and}
  \bibinfo{author}{\bibfnamefont{A.}~\bibnamefont{Smerzi}}:
  \bibinfo{journal}{Phys. Rev. Lett.} \textbf{\bibinfo{volume}{93}} (\bibinfo{year}{2004})
  \bibinfo{pages}{110401}.

\bibitem{Ott04}
  \bibinfo{author}{\bibfnamefont{E.}~\bibnamefont{de Mirandes}},
  \bibinfo{author}{\bibfnamefont{F.}~\bibnamefont{Ferlaino}},
  \bibinfo{author}{\bibfnamefont{G.}~\bibnamefont{Roati}},
  \bibinfo{author}{\bibfnamefont{G.}~\bibnamefont{Modugno}}, \bibnamefont{and}
  \bibinfo{author}{\bibfnamefont{M.}~\bibnamefont{Inguscio}}:
  \bibinfo{journal}{Phys. Rev. Lett.} \textbf{\bibinfo{volume}{92}} (\bibinfo{year}{2004})
  \bibinfo{pages}{160601}.

\bibitem{Gib95}
\bibinfo{author}{\bibfnamefont{K.}~\bibnamefont{Gibble}} \bibnamefont{and}
  \bibinfo{author}{\bibfnamefont{B.~J.} \bibnamefont{Verhaar}}:
  \bibinfo{journal}{Phys. Rev. A} \textbf{\bibinfo{volume}{52}} (\bibinfo{year}{1995})
  \bibinfo{pages}{3370}.

\bibitem{Kat03}
\bibinfo{author}{\bibfnamefont{H.}~\bibnamefont{Katori}},
  \bibinfo{author}{\bibfnamefont{M.}~\bibnamefont{Takamoto}},
  \bibinfo{author}{\bibfnamefont{V.~G.} \bibnamefont{Pal\char39{}chikov}},
  \bibnamefont{and} \bibinfo{author}{\bibfnamefont{V.~D.}
  \bibnamefont{Ovsiannikov}}: \bibinfo{journal}{Phys. Rev. Lett.}
  \textbf{\bibinfo{volume}{91}} (\bibinfo{year}{2003}) \bibinfo{pages}{173005}.

\bibitem{Blo92}
\bibinfo{author}{\bibfnamefont{C.~A.} \bibnamefont{Blockley}},
  \bibinfo{author}{\bibfnamefont{D.~F.} \bibnamefont{Walls}}, \bibnamefont{and}
  \bibinfo{author}{\bibfnamefont{H.}~\bibnamefont{Risken}}:
  \bibinfo{journal}{Europhys. Lett.} \textbf{\bibinfo{volume}{17}} (\bibinfo{year}{1992})
  \bibinfo{pages}{509}.

\bibitem{Cir94}
\bibinfo{author}{\bibfnamefont{J.~I.} \bibnamefont{Cirac}},
  \bibinfo{author}{\bibfnamefont{R.}~\bibnamefont{Blatt}},
  \bibinfo{author}{\bibfnamefont{A.~S.} \bibnamefont{Parkins}},
  \bibnamefont{and} \bibinfo{author}{\bibfnamefont{P.}~\bibnamefont{Zoller}}:
  \bibinfo{journal}{Phys. Rev. A} \textbf{\bibinfo{volume}{49}} (\bibinfo{year}{1994})
  \bibinfo{pages}{1202}.

\bibitem{Tak05}
\bibinfo{author}{\bibfnamefont{M.}~\bibnamefont{Takamoto}},
  \bibinfo{author}{\bibfnamefont{F.-L.} \bibnamefont{Hong}},
  \bibinfo{author}{\bibfnamefont{R.}~\bibnamefont{Higashi}}, \bibnamefont{and}
  \bibinfo{author}{\bibfnamefont{H.}~\bibnamefont{Katori}}:
  \bibinfo{journal}{Nature} \textbf{\bibinfo{volume}{435}} (\bibinfo{year}{2005})
  \bibinfo{pages}{321}.

\bibitem{Per02}
\bibinfo{author}{\bibfnamefont{P.~D.} \bibnamefont{Santos}},
  \bibinfo{author}{\bibfnamefont{H.} \bibnamefont{Marion}},
  \bibinfo{author}{\bibfnamefont{S.} \bibnamefont{Bize}},
  \bibinfo{author}{\bibfnamefont{Y.} \bibnamefont{Sortais}},
  \bibinfo{author}{\bibfnamefont{A.} \bibnamefont{Clairon}}, \bibnamefont{and}
  \bibinfo{author}{\bibfnamefont{C.} \bibnamefont{Salomon}}:
  \bibinfo{journal}{Phys. Rev. Lett.}
  \textbf{\bibinfo{volume}{89}} (\bibinfo{year}{2002}) \bibinfo{pages}{233004}.

\bibitem{Wil02}
\bibinfo{author}{\bibfnamefont{G.}~\bibnamefont{Wilpers}},
  \bibinfo{author}{\bibfnamefont{T.}~\bibnamefont{Binnewies}},
  \bibinfo{author}{\bibfnamefont{C.}~\bibnamefont{Degenhardt}},
  \bibinfo{author}{\bibfnamefont{U.}~\bibnamefont{Sterr}},
  \bibinfo{author}{\bibfnamefont{J.}~\bibnamefont{Helmcke}}, \bibnamefont{and}
  \bibinfo{author}{\bibfnamefont{F.}~\bibnamefont{Riehle}}:
  \bibinfo{journal}{Phys. Rev. Lett.} \textbf{\bibinfo{volume}{89}} (\bibinfo{year}{2002})
  \bibinfo{pages}{230801}.

\bibitem{Dem01}
\bibinfo{author}{\bibfnamefont{B.}~\bibnamefont{DeMarco}},
  \bibinfo{author}{\bibfnamefont{S.~B.} \bibnamefont{Papp}}, \bibnamefont{and}
  \bibinfo{author}{\bibfnamefont{D.~S.} \bibnamefont{Jin}}:
  \bibinfo{journal}{Phys. Rev. Lett.} \textbf{\bibinfo{volume}{86}} (\bibinfo{year}{2001})
  \bibinfo{pages}{5409}.

\bibitem{Gup03}
\bibinfo{author}{\bibfnamefont{S.}~\bibnamefont{Gupta}},
  \bibinfo{author}{\bibfnamefont{Z.}~\bibnamefont{Hadzibabic}},
  \bibinfo{author}{\bibfnamefont{M.~W.}~\bibnamefont{Zwierlein}},
  \bibinfo{author}{\bibfnamefont{C.~A.}~\bibnamefont{Stan}},
  \bibinfo{author}{\bibfnamefont{K.}~\bibnamefont{Dieckmann}},
  \bibinfo{author}{\bibfnamefont{C.~H.}~\bibnamefont{Schunck}},
  \bibinfo{author}{\bibfnamefont{E.~G.~M.}~\bibnamefont{van Kempen}},
  \bibinfo{author}{\bibfnamefont{B.~J.}~\bibnamefont{Verhaar}}, \bibnamefont{and}
  \bibinfo{author}{\bibfnamefont{W.}~\bibnamefont{Ketterle}}:
  \bibinfo{journal}{Science} \textbf{\bibinfo{volume}{300}} (\bibinfo{year}{2003})
  \bibinfo{pages}{1723}.

\bibitem{Tak06}
\bibinfo{author}{\bibfnamefont{M.}~\bibnamefont{Takamoto}},
  \bibinfo{author}{\bibfnamefont{F.~L.}~\bibnamefont{Hong}},
  \bibinfo{author}{\bibfnamefont{R.}~\bibnamefont{Higashi}},
   \bibinfo{author}{\bibfnamefont{Y.}~\bibnamefont{Fujii}},
   \bibinfo{author}{\bibfnamefont{M.}~\bibnamefont{Imae}}, \bibnamefont{and}
   \bibinfo{author}{\bibfnamefont{H.}~\bibnamefont{Katori}}:
  \bibinfo{journal}{J. Phys. Soc. Jpn.}
  \textbf{\bibinfo{volume}{75}} (\bibinfo{year}{2006}) \bibinfo{pages}{104302}.

\bibitem{Lud08}
\bibinfo{author}{\bibfnamefont{A.~D.} \bibnamefont{Ludlow}},
\bibinfo{author}{\bibfnamefont{T.} \bibnamefont{Zelevinsky}},
\bibinfo{author}{\bibfnamefont{G.~K.} \bibnamefont{Campbell}},
\bibinfo{author}{\bibfnamefont{S.} \bibnamefont{Blatt}},
\bibinfo{author}{\bibfnamefont{M.~M.} \bibnamefont{Boyd}},
\bibinfo{author}{\bibfnamefont{M.~H.~G.} \bibnamefont{de Miranda}},
\bibinfo{author}{\bibfnamefont{M.~J.} \bibnamefont{Martin}},
\bibinfo{author}{\bibfnamefont{J.~W.} \bibnamefont{Thomsen}},
\bibinfo{author}{\bibfnamefont{S.~M.} \bibnamefont{Foreman}},
\bibinfo{author}{\bibfnamefont{J.} \bibnamefont{Ye}},
\bibinfo{author}{\bibfnamefont{T.~M.} \bibnamefont{Fortier}},
\bibinfo{author}{\bibfnamefont{J.~E.} \bibnamefont{Stalnaker}},
\bibinfo{author}{\bibfnamefont{S.~A.} \bibnamefont{Diddams}},
\bibinfo{author}{\bibfnamefont{Y.} \bibnamefont{Le Coq}},
\bibinfo{author}{\bibfnamefont{Z.~W.} \bibnamefont{Barber}},
\bibinfo{author}{\bibfnamefont{N.} \bibnamefont{Poli}},
\bibinfo{author}{\bibfnamefont{N.~D.} \bibnamefont{Lemke}},
\bibinfo{author}{\bibfnamefont{K.~M.} \bibnamefont{Beck}}, \bibnamefont{and}
\bibinfo{author}{\bibfnamefont{C.~W.} \bibnamefont{Oates}}:
  \bibinfo{journal}{Science}
  \textbf{\bibinfo{volume}{319}} (\bibinfo{year}{2008}) \bibinfo{pages}{1805}.

\bibitem{Lei03}
\bibinfo{author}{\bibfnamefont{D.}~\bibnamefont{Leibfried}},
  \bibinfo{author}{\bibfnamefont{R.}~\bibnamefont{Blatt}},
  \bibinfo{author}{\bibfnamefont{C.}~\bibnamefont{Monroe}}, \bibnamefont{and}
  \bibinfo{author}{\bibfnamefont{D.}~\bibnamefont{Wineland}}:
  \bibinfo{journal}{Rev. Mod. Phys.} \textbf{\bibinfo{volume}{75}} (\bibinfo{year}{2003})
  \bibinfo{pages}{281}.

\bibitem{Muk03}
\bibinfo{author}{\bibfnamefont{T.}~\bibnamefont{Mukaiyama}},
  \bibinfo{author}{\bibfnamefont{H.}~\bibnamefont{Katori}},
  \bibinfo{author}{\bibfnamefont{T.}~\bibnamefont{Ido}},
  \bibinfo{author}{\bibfnamefont{Y.}~\bibnamefont{Li}}, \bibnamefont{and}
  \bibinfo{author}{\bibfnamefont{M.}~\bibnamefont{Kuwata-Gonokami}}:
  \bibinfo{journal}{Phys. Rev. Lett.} \textbf{\bibinfo{volume}{90}} (\bibinfo{year}{2003})
  \bibinfo{pages}{113002}.

\bibitem{ft1}
  The overlap may be improved up to 0.13~mrad for the current experiment, which is limited by a wavefront curvature of  Gaussian beams.

\bibitem{ft2}
  The Bloch bandwidths were expected to be $\gamma_0'/2\pi=0.05$~Hz \cite{Sia08}  and $\gamma_1'/2\pi \approx 200$~Hz for $\theta=10$~mrad.

\bibitem{Kat99}
\bibinfo{author}{\bibfnamefont{H.}~\bibnamefont{Katori}},
  \bibinfo{author}{\bibfnamefont{T.}~\bibnamefont{Ido}}, \bibnamefont{and}
  \bibinfo{author}{\bibfnamefont{M.}~\bibnamefont{Kuwata-Gonokami}}:
  \bibinfo{journal}{J. Phys. Soc. Jpn.} \textbf{\bibinfo{volume}{68}} (\bibinfo{year}{1999})
  \bibinfo{pages}{2479}.

\bibitem{Kel88}
\bibinfo{author}{\bibfnamefont{J.~F.} \bibnamefont{Kelly}},
  \bibinfo{author}{\bibfnamefont{M.}~\bibnamefont{Harris}}, \bibnamefont{and}
  \bibinfo{author}{\bibfnamefont{A.}~\bibnamefont{Gallagher}}:
  \bibinfo{journal}{Phys. Rev. A} \textbf{\bibinfo{volume}{38}} (\bibinfo{year}{1988})
  \bibinfo{pages}{1225}.

\bibitem{Lem05}
\bibinfo{author}{\bibfnamefont{P.}~\bibnamefont{Lemonde}} \bibnamefont{and}
  \bibinfo{author}{\bibfnamefont{P.}~\bibnamefont{Wolf}}:
  \bibinfo{journal}{Phys. Rev. A} \textbf{\bibinfo{volume}{72}} (\bibinfo{year}{2005})
  \bibinfo{pages}{033409}.

\bibitem{Sia08}
\bibinfo{author}{\bibfnamefont{C.}~\bibnamefont{Sias}},
  \bibinfo{author}{\bibfnamefont{H.}~\bibnamefont{Lignier}},
  \bibinfo{author}{\bibfnamefont{Y.~P.}~\bibnamefont{Singh}},
  \bibinfo{author}{\bibfnamefont{A.}~\bibnamefont{Zenesini}},
  \bibinfo{author}{\bibfnamefont{D.}~\bibnamefont{Ciampini}},
  \bibinfo{author}{\bibfnamefont{O.}~\bibnamefont{Morsch}}, \bibnamefont{and}
  \bibinfo{author}{\bibfnamefont{E.}~\bibnamefont{Arimondo}}:
  \bibinfo{journal}{Phys. Rev. Lett.} \textbf{\bibinfo{volume}{100}} (\bibinfo{year}{2008})
  \bibinfo{pages}{040404}.

\bibitem{San04}
\bibinfo{author}{\bibfnamefont{R.}~\bibnamefont{Santra}},
  \bibinfo{author}{\bibfnamefont{K.~V.} \bibnamefont{Christ}}, \bibnamefont{and} 
  \bibinfo{author}{\bibfnamefont{C.~H.} \bibnamefont{Greene}}: 
  \bibinfo{journal}{Phys. Rev. A} \textbf{\bibinfo{volume}{69}} (\bibinfo{year}{2004})
  \bibinfo{pages}{042510}.

\end{thebibliography}
\end{document}